\documentclass[prb,aps,twocolumn,epsf,showpacs,preprintnumbers,amsmath,amssymb]{revtex4-1}
\usepackage{graphicx}% Include figure files
\usepackage{dcolumn}% Align table columns on decimal point
\usepackage{bm}% bold math
\usepackage{ulem}%strikeout
\begin{document}
%\baselineskip=16pt

%\preprint{working draft}

\title{ Andreev-Lifshitz Hydrodynamics Applied to an Ordinary Solid under Pressure} 
\author{ Matthew R. Sears} 
\author{ Wayne M. Saslow} 
\email{wsaslow@tamu.edu}
\affiliation{ Department of Physics, Texas A\&M University, College Station, 
TX 77843-4242}
\date{\today}

\begin{abstract}

We have applied the Andreev-Lifshitz hydrodynamic theory of supersolids to an ordinary solid.    This theory includes an internal pressure $P$, distinct from the applied pressure $P_a$ and the stress tensor $\lambda_{ik}$.  Under uniform static $P_{a}$, we have $\lambda_{ik}=(P-P_{a})\delta_{ik}$.  For $P_{a}\ne0$, Maxwell relations imply that $P\sim P_{a}^{2}$.  The theory also permits vacancy diffusion but treats vacancies as conserved. 
It gives three sets of propagating elastic modes; it also gives two diffusive modes, one largely of entropy density and one largely of vacancy density (or, more generally, defect density).    
For the vacancy diffusion mode (or, equivalently, the lattice diffusion mode) the vacancies behave like a fluid within the solid, with the deviations of internal pressure associated with density changes nearly canceling the deviations of stress associated with strain.  We briefly consider pressurization experiments in solid $^4$He at low temperatures in light of this lattice diffusion mode, which for small $P_{a}$ has diffusion constant $D_{L}\sim P_{a}^{2}$.  The general principles of the theory -- that both volume and strain should be included as thermodynamic variables, with the result that both $P$ and $\lambda_{ik}$ appear -- should apply to all solids under pressure, especially near the solid-liquid transition.  
The lattice diffusion mode provides an additional degree of freedom that may permit surfaces with different surface treatments to generate different responses in the bulk.

\end{abstract}

\pacs{67.80.B-, 67.80.bd, 05.70.Ln, 63.10.+a}

%72.20.-i, 72.20 Jv, 72.40.+w, 73.50.Pz, , 61.72.Ww, 61.72.Ss ????}

\maketitle

\section{Introduction}

Since the late 1960's there have been theoretical suggestions that solids might display flow behavior similar to what is found in superfluids.\cite{AL69,Thouless69,Chester70,Leggett70}  For that reason there has been a great deal of interest in solid $^4$He as a candidate {\it supersolid}.\cite{BalCau08}  The first experimental indication of superflow was the appearance of a non-classical moment of inertia (NCRI), first observed by Chan's group, since confirmed by many other laboratories, and strongly linked to disorder.\cite{KC1,KC2,RR1,Shira1,Koj1,Penzev,RR2,RR3,Lin07,ClarkWestChan07}  In addition, the shear modulus shows anomalous behavior,\cite{DayBeamishNature07} although not enough to explain the NCRI experiments.\cite{ChanScience08}  Non-NCRI superflow has  been searched for but not observed.\cite{Day06}  We also note recent experiments that argue against any supersolid signature above approximately 55 mK.\cite{ShevDayBeam10}  Further works casting doubt on supersolidity are a study of bcc $^4$He that shows unusual NCRI behavior at higher temperatures,\cite{SatanPEPolturak} and a study showing that the NCRI behavior due to plasticity has different properties than due to quenching.\cite{Reppy10} 

In a recent experiment on a pancake-shaped sample, a capacitance gauge monitored the pressure as a function of temperature $T$.\cite{RitRep09}  Samples were produced by both the slow-cooling blocked capillary method and by the more rapid quench-cooling method, which gives more disordered samples.  In one set of measurements the sample was quench-cooled below 1~K in 144~s, during which time the pressure decreased.  This is perhaps an indication that vacancies, formed during the quench, were leaving the sample.  For a blocked-capillary sample the temperature was lowered below 500~mK while the pressure was monitored {\it vs} $T$.  The sample was then annealed at 1.65~K, where the pressure increased, perhaps an indication that vacancies now were entering the sample.  A second cooldown yielded, by a reduced $T^{2}$ term in the pressure, an indication that the sample was less disordered, but that disorder remained.  Even at a constant temperature of 19~mK the pressure continued to relax, which is consistent with vacancies equilibrating.  The fact that the observed relaxation times do not saturate at the temperatures studied indicates that the temperature is not yet low enough that quantum relaxation processes dominate thermal relaxation processes.  

This experiment can perhaps be interpreted under the assumption that the system is {\it not} supersolid.  We have therefore undertaken a theoretical study of the macroscopic flow properties of a one-component ordinary solid.  Our basis is the theory of Andreev and Lifshitz (AL) for the macroscopic behavior of a supersolid.  They included volume $V$ as an extensive variable, in addition to $W_{ik}\equiv V w_{ik}$, where $w_{ik}$ is the non-symmetrized strain.  This permitted them to continuously go to the superfluid limit as $w_{ik}$ becomes irrelevant.  The point of the present work is that on eliminating the superfluid variables, the theory should apply to an ordinary solid.\cite{AL69}  We employ a variation on the notation of Ref.~\onlinecite{Saslow77}, which gives a more explicit derivation of the equations of motion and extends Ref.~\onlinecite{AL69} to include nonlinear terms.\cite{SaslowNote,Liu}  Note also the theory of Fleming and Cohen\cite{FC76} for an ordinary solid, which gives equations with a similar structure,  and similar modes, but uses a very different notation (and does not consider an applied pressure $P_{a}$).  Both Ref.~\onlinecite{AL69} and Ref.~\onlinecite{FC76} implicitly assume that uniform vacancy number-changing bulk processes are negligible, and neglect interstitials and impurities.  Recently Yoo and Dorsey\cite{YooDorsey} considered the effect of a lattice diffusion mode on light scattering by a supersolid, but also briefly considering an ordinary solid.
 
As noted by Martin, Parodi, and Pershan,\cite{MartinParodi} the normal system has eight degrees of freedom, given by two scalar thermodynamic quantities (which can be taken to be the mass density $\rho$ and the entropy density $s$) and two vector quantities: the lattice vector $u_{i}$ and the velocity $v_{i}$ associated with the momentum density $g_{i}=\rho v_{i}$.  (With $m_{4}$ the atomic mass and $n$ the number density of $^4$He atoms, we have $\rho=m_{4}n$.)  As a consequence there are eight normal modes.  For a uniform infinite system these modes are three pairs of propagating elastic waves and two diffusive modes, one primarily of the temperature $T$ and the other primarily of $\partial_{i} u_{i}$.  In the absence of lattice defects, for a variation $\delta u_{i}$ the relationship 
\begin{align}
\partial_{i} (\delta u_{i}) \approx-\delta\rho/\rho
\label{uTOrhoNODEFECT}
\end{align}
holds, giving the system one fewer degree of freedom, and thus one fewer mode.  One can think of this missing mode, associated with the dynamical violation of \eqref{uTOrhoNODEFECT}, as being associated with vacancies, as noted in Ref.~\onlinecite{MartinParodi}. 

The present work obtains the diffusion constant and the physical properties of this diffusive mode, for both zero and non-zero $P_a$.  ($^4$He must be under $P_{a}\approx$ 25 atmospheres to solidify.)  We find that the physical character of the mode is that it involves essentially zero stress deviation, because the fluid-like stress (associated with changes in mass density) nearly cancels the solid-like stress (associated with changes in strain). 

Allowing vacancies to move permits mass change without lattice motion.\cite{BardeenHerring}  This allows one to take the fluid limit of zero crystallinity, and study the evolution of the sound velocity as the system evolves from the perfect solid to perfect liquid.  By perfect solid we mean one with no defects and a one-to-one relationship between lattice points and atoms; by perfect liquid we mean one with no lattice structure or, equivalently, one with no sensitivity to an imaginary lattice structure.  A gel has properties of both, but is multi-component.\cite{DLJohnson82} 
 
Section~\ref{Dynamics} gives the form of AL supersolid theory when restricted to a normal solid, including the possibility of lattice defects.  Although we specifically have vacancies in mind, $^3$He impurities could be accounted for if its density were included as an additional thermodynamic variable, which would require extension of the AL theory.  Note also the case of (two-component) superionic conductors, which includes certain high-temperature alkali halides, where the larger halide ions remain in a lattice but the lattice of the smaller alkali ions ``melts.''  
Section~\ref{CrystalPressure} discusses elasticity and internal pressure for a crystal under static and uniform applied pressure $P_a$, and calculates internal pressure $P$ and strain.  We find that $P \sim P_a^2$, so that for small $P_{a}$ the effect of $P$ is very small; see eq.~\eqref{P2}.  For small $P_{a}$ the strain is largely linear in $P_{a}$, as expected, but there is a $P_{a}^{2}$ correction.  
Section~\ref{NormalModesSec} derives the normal modes for the ordinary solid. Section~\ref{ModeGeneration} considers how such modes can be generated (including the possible effect of different surface treatments), and applies the theory to the pressurization experiments.\cite{RitRep09}   Section~\ref{Summary} provides a summary and our conclusions.  Appendix~\ref{ALAppendix} gives the thermodynamics and dynamics of the AL theory for the supersolid.   Appendix~\ref{DerivativeAppendix} calculates some thermodynamic derivatives that appear in the normal modes in terms of $P_a$.  %Appendix~\ref{Kappendix} briefly discusses the unconventional nature of some relevant derivatives.

\section{Andreev-Lifshitz Normal Solid with Defects}
\label{Dynamics}
We employ the primary quantities energy density $\epsilon$, lattice displacement $u_{i}$, and non-symmetrized strain 
\begin{align}
w_{ik}=\partial_{i}u_{k}.
\label{wDEF}
\end{align}

We consider a normal solid by  setting $\rho_{s} =0$, $\rho_{n} = \rho$, ${\vec{v}}_n = \vec{v}$, and eliminating the superfluid equation from the equations for the supersolid (given in Appendix~\ref{ALAppendix}).  

\subsection{Thermodynamics}
The appropriate thermodynamic equations are 
\begin{eqnarray}
d\epsilon&=&Tds+\lambda_{ik}dw_{ik}+ \mu d\rho +\vec{v}\cdot d\vec{g}, \label{n-depsilon}\\ 
\epsilon&=&-P+Ts+\lambda_{ik}w_{ik}+\mu \rho +\vec{v}\cdot\vec{g}, \label{n-epsilon}\\
0&=&-dP+sdT+w_{ik}d\lambda_{ik}+\rho d\mu +\vec{g}\cdot d\vec{v}. \label{n-GibbsDuhem}
\label{thermoNormal}
\end{eqnarray}
Here $\lambda_{ik}$ of AL is an elastic tensor density (with units of pressure $P$), and $\mu$ is the chemical potential (with units of velocity squared); $\lambda_{ik}$ is the same as $\sigma_{ik}$ of Ref.~\onlinecite{LLElasticity}. 

\subsection{Dynamics}
The appropriate linearized equations of motion for the independent variables $s$, $u_{i}$, $\rho$, and $v_i$ are
\begin{eqnarray}
\partial_{t}s+\partial_{i}f_{i}&=&0, \label{n-sdot}\\
\partial_{t}u_{i}&=&U_{i}, \label{n-udot}\\
\partial_{t}\rho+\partial_{i}g_{i}&=&0, \label{n-rhodot}\\
\partial_{t}g_{i}+\partial_{k}\Pi_{ik}&=&0, \label{n-gdot}
\label{eqnsmotNorm}
\end{eqnarray}
where the fluxes $f_{i}$ (of entropy), $\Pi_{ik}$ (of momentum), $g_{i}$ (of mass), and the ``source'' $U_{i}$ (terminology introduced here) are given by
\begin{eqnarray}
f_{i}&=&sv_{i}-\frac{\kappa_{ij}}{T}\partial_{j}T-\frac{\alpha_{ij}}{T}\partial_{l}\lambda_{lj}, \label{n-f}\\
U_{i}&=&v_{i}+\frac{\alpha_{ij}}{T}\partial_{j}T+\beta_{ij}\partial_{l}\lambda_{lj}, \label{n-U}\\
\Pi_{ik}&=&(P\delta_{ik}-\lambda_{ik})-\eta_{iklm}\partial_{m}v_{l}, \label{n-Pi}\\
g_{i}&=&\rho v_{i}. \label{n-g}
\label{fluxsource}
\end{eqnarray}
AL use $\sigma_{ik} \approx - \Pi_{ik}$.\cite{NotationFootnote} 
The term in \eqref{n-U} proportional to $\beta_{ij}$ allows the lattice velocity $\dot{u}_{i}$ to differ from the velocity $v_{i}$ associated with mass flow.  It leads, as we show, to a lattice diffusion mode for which $\dot{u}_{i}\ne v_{i}$ and neither is zero.  %Our notation for the dissipative coefficients follows AL. 

Both the $\alpha_{ij}$ and $\beta_{ij}$ terms can be rewritten as flux terms.  Linearizing about equilibrium,  with primes denoting deviations from equilibrium, yields
\begin{align}
\partial_t u_i + \partial_j S_{ij} = v_i',
\label{n-udotREWRITE}
\end{align}
where
\begin{align}
S_{ij} \equiv -\frac{\alpha_{ij}}{T} T' - \beta_{il} \lambda_{jl}',
\label{Sij}
\end{align}
In \eqref{n-udotREWRITE}, $v_i$ can be thought of as a ``lattice source'', and $S_{ij}$ as a ``lattice flux.'' The $\beta_{il}$ term gives, in  principle, anisotropic vacancy diffusion.

Recall that a diffusion constant $D$ is proportional to a characteristic velocity times a characteristic mean-free path, so it has units of m$^{2}$/sec.  In terms of a $D$, the dissipative coefficients have the following units: $\kappa_{ij}$ has units of $s$ times $D$; $\alpha_{ij}$ has units of $D$; $\beta_{ij}$ has units of inverse pressure times $D$; and $\eta_{iklm}$ has units of $\rho$ times $D$.  

\section{Crystal Under Pressure}
\label{CrystalPressure}

\subsection{Internal Pressure and Elasticity} 
\label{PressureElasticity}
The momentum conservation equation \eqref{n-gdot} implicitly contains the term $\lambda_{ik}-P\delta_{ik}$, which determines the force on the surface of the solid.  
An internal pressure $P$ does not appear in the thermodynamics of Ref.~\onlinecite{LLElasticity}, which does not consider either a lattice under applied pressure or the presence of defects.  However, the extensive energy $E=\epsilon V$, which depends on the extensive variables $(S,V,N,W_{ik} \equiv Vw_{ik},V\vec{g})$, has second derivatives that satisfy the Maxwell relation 
\begin{equation}
-\frac{\partial P}{\partial W_{ik}}=\frac{\partial \lambda_{ik}}{\partial V}, 
\label{Maxwell1}
\end{equation}
where the appropriate variables are held constant.  For solid $^4$He under an applied pressure $P_{a}$, this makes $P$ non-zero.  

In principle we may let $E$ depend on the number of vacancies $N_{V}$, with associated ``chemical potential'' $\phi_{V}=\partial E/\partial N_{V}$ (with units of energy, rather than velocity squared).  Then the additional Maxwell relation 
\begin{equation}
-\frac{\partial P}{\partial N_{V}}=\frac{\partial\phi_{V}}{\partial V} 
\label{Maxwell2}
\end{equation}
follows, with the appropriate variables held constant.  If the vacancies are not in equilibrium (i.e., $\phi_{V}\neq 0$), this also makes $P$ non-zero.  The general results of the present work (e.g., a nonzero lattice diffusion constant) can thus be made applicable to a solid not under $P_a$ but having vacancies out of local thermal equilibrium.  Terms found here to depend on $P_a$ may in that case depend on the difference between actual concentration of vacancies and the equilibrium concentration of vacancies. %[I'm not sure whether to make direct reference here to Yoo-Dorsey or to Fleming, Martin, etc.  Seems strange to call them out, and hypocritical considering our statement in our reply that ``we do not want to make statements about what others did NOT do" (Paraphrased).] 
However, we expect a $P_a$ of $\sim 25$~atm to dominate the effect of vacancies, and thus we neglect their effect on $P$.  Although Refs.~\onlinecite{AL69}, \onlinecite{FC76} and \onlinecite{MartinParodi} introduce the internal pressure $P$, they do not calculate $P$ or its thermodynamic derivatives.  Ref.~\onlinecite{AL69} and the present work neglect the possibility of interstitial atoms.\cite{DefectFootnote}  For a reference that considers interstitials, see Ref.~\onlinecite{Zippelius}.

As employed by Ref.~\onlinecite{AL69}, this pressure term, in contrast to $\lambda_{ik}$ alone (Ref.~\onlinecite{LLElasticity} does {\it not} include $P$), permits one to continuously approach the superfluid limit, when the lattice disappears.  In the present case, it permits one to continuously approach the ordinary liquid limit.

The consequences of a nonzero $P$ include, but are not limited to, a mode where vacancies are permitted to diffuse.  Thermodynamic derivatives of $P$ are essential for defect diffusion, and also affect the elastic modes.  Moreover, they are needed to obtain the pure liquid limit for longitudinal sound on letting the crystallinity go to zero.  We first use a Maxwell relation to find an explicit expression for $P$ as a function of strain.

\subsection{Internal Pressure $P$}
Since holding $(V, N)$ constant is equivalent to holding $(V,\rho)$ constant, and similarly for $(S,N)$ and $(\sigma=s/\rho,N)$, we use these sets interchangeably.  We rewrite \eqref{Maxwell1} as
\begin{align}
-\left. \frac{\partial P}{\partial W_{ik}} \right|_{V,S,N} = -\frac{1}{V} \left. \frac{\partial P}{\partial w_{ik}} \right|_{V,\sigma,\rho} = \left. \frac{\partial \lambda_{ik}}{\partial V} \right|_{W_{ik},S,N}.
\label{dPdWmaxwell}
\end{align}
For constant $W_{ik}$ we have
\begin{align}
0 = dW_{ik} = w_{ik} dV + V dw_{ik},
\end{align}
so that 
\begin{align}
\left. \frac{dw_{ik}}{\partial V} \right|_{W_{ik} ,S , N} = -\frac{w_{ik}}{V} .
\end{align}
Then
\begin{align}
&\left. \frac{\partial \lambda_{ik}}{\partial V} \right|_{W_{ik},S,N}  =   \left. \frac{\partial \lambda_{ij}}{\partial V} \right|_{w_{ik},\sigma , N} - \frac{w_{jl}}{V} \left. \frac{\partial \lambda_{ik}}{\partial w_{jl}} \right|_{V,\sigma , \rho} ,
\end{align}
and \eqref{dPdWmaxwell} gives
\begin{align}
\left. \frac{\partial P}{\partial w_{ik}} \right|_{V,\sigma,\rho}=& - V \left. \frac{\partial \lambda_{ik}}{\partial V} \right|_{W_{ik},S,N} \notag\\ 
=& - V\left. \frac{\partial \lambda_{ij}}{\partial V} \right|_{w_{ik},\sigma,N} + {w_{jl}} \left. \frac{\partial \lambda_{ik}}{\partial w_{jl}} \right|_{V,\sigma,\rho} .
\label{dPdwTOdlambda}
\end{align}

We employ Ref.~\onlinecite{LLElasticity} for the elasticity tensor $\lambda_{ik}$ in an isotropic solid.  Using superscript $(0)$ to denote the equilibrium value of $\lambda_{ik}$ and the strain $w_{ik}$, we have
\begin{align}
\lambda_{ik}^{(0)} = \left(K - \frac{2}{3}\mu_V \right) \delta_{ik} w_{ll}^{(0)} + \mu_V \left(w_{ik}^{(0)} + w_{ki}^{(0)}\right),
\label{lambdaNEW}
\end{align}
where $K$ and $\mu_V$ are the bulk and shear moduli, and both $\lambda_{ik}^{(0)}$ and $w_{ik}^{(0)}$ are to be determined under a given applied pressure $P_{a}$.  Eq.~\eqref{dPdwTOdlambda} then gives
\begin{align}
\left. \frac{\partial P}{\partial w_{ik}} \right|_{V,\sigma ,\rho}=& \left(K^* - \frac{2}{3} \mu_V^* \right)\delta_{ik} w_{ll}^{(0)} + \mu_V^* \left(w_{ik}^{(0)} + w_{ki}^{(0)}\right),
\label{dPdw1}
\end{align}
where
\begin{align}
K^* = K - V \left. \frac{\partial K}{\partial V} \right|_{w_{ik},\sigma , N}, \quad \mu_V^* = \mu_V - V \left. \frac{\partial \mu_V}{\partial V}\right|_{w_{ik},\sigma ,N}.
\label{Kmustar}
\end{align}
%Appendix~\ref{Kappendix} discusses the unconventional derivative of $K$ that appears in \eqref{Kmustar}.

Under uniform $P_{a}$ we expect an isotropic response, so 
\begin{align}
w_{ik}^{(0)} = \frac{\delta_{ik}}{3} w_{ll}^{(0)}.
\label{wik0SYM} 
\end{align}
Then \eqref{dPdw1} becomes
\begin{align}
\left. \frac{\partial P}{\partial w_{ik}} \right|_{V,\sigma,\rho}=& K^* \delta_{ik} w_{ll}^{(0)} .
\label{dPdw1.5}
\end{align}
Integration of \eqref{dPdw1.5} with respect to $w_{ik}$ gives the part of the internal pressure dependent on the strain to be
\begin{align}
P = \frac{1}{2} K^* \left(w_{ll}^{(0)} \right)^2 ,
\end{align}
where we take the integration constant to be zero.\cite{NonzeroPFootnote}  For ${w_{11}^{(0)}} = {w_{22}^{(0)}} = {w_{33}^{(0)}}$,
we then have 
\begin{align}
P = \frac{9}{2} K^* {w_{11}^{(0)}}^2.
\label{P}
\end{align}
%\pagebreak

This result applies to the case of a strongly crystalline material.  In the opposite limit where the crystallinity disappears and the particles are weakly interacting, part of $P$ would be given by the ideal gas law.  

\subsection{Strain $w_{ik}$}
As discussed above, under an applied pressure the force on the surface of a solid is
\begin{align}
\lambda_{ik}^{(0)} - P \delta_{ik} = - P_{a} \delta_{ik}.
\end{align}
Taking the trace yields
\begin{align}
\frac{\lambda_{ll}^{(0)}}{3} - P = - P_{a}.
\label{stress0}
\end{align}
Substitution from \eqref{lambdaNEW} and \eqref{P} gives
\begin{align}
3 K {w_{11}^{(0)}} - \frac{9}{2} K^* {w_{11}^{(0)}}^2 = -P_a.
\label{lambdaTOwNEW}
\end{align}
Since an applied pressure should cause a negative strain, only the solution for ${w_{11}^{(0)}}<0$ is physical. 

For solid $^4$He, we expect both ${w_{11}^{(0)}}$ and $P_a/K$ to be small.  The solution of \eqref{lambdaTOwNEW} to second order in $P_{a}$ is
\begin{align}
{w_{11}^{(0)}} \approx -\frac{P_a}{3K} +  \frac{P_a^2 K^*}{6 K^3}.
\label{w11expand}
\end{align}
The first term is what one would get on neglecting $P$ in \eqref{stress0}.  To second order in $P_a/K$, eq.~\eqref{P} then gives
\begin{align}
P =  K^* \frac{P_a^2}{2 K^2},
\label{P2}
\end{align}
a result that appears to be new.  Further, $\lambda_{ik}^{(0)}=\delta_{ik}\lambda_{11}^{(0)}$, where
\begin{align}
\lambda_{11}^{(0)} = -P_{a}+  K^* \frac{P_a^2}{2 K^2}.
\label{lambda0}
\end{align}
The first term in $\lambda_{11}^{(0)}$ is what one obtains on neglecting $P$ in \eqref{stress0}, and in agreement with Ref.~\onlinecite{LLElasticity}.

\section{Normal Modes of Andreev-Lifshitz Normal Solid with Defects}
\label{NormalModesSec}
As noted earlier, this system has eight variables: $s$, $\rho$, $g_{i}$ and $u_{i}$.  Disturbances from equilibrium will be denoted by primes, so we use $s'$, $\rho'$, $g'_{i}\approx\rho v'_{i}$, and $u'_{i}$.  There are correspondingly eight normal modes.  For an infinite system we assume a disturbance of the form $\exp[i(\vec{k}\cdot \vec{r}-\omega t)]$, where the real wavevector $\vec{k}$ is considered to be known, but $\omega$ is unknown.  For the disturbance to decay in time, $Im(\omega)<0$.  Six modes come in three degenerate pairs, with $g_{i}'$ and $u_{i}'$ strongly coupled, and correspond to ordinary elasticity.  The other two modes are diffusive, with temperature diffusion nearly decoupled from lattice diffusion.  To ensure this decoupling we set the (off-diagonal) temperature-lattice transport coefficient $\alpha_{ij}=0$, and set the distinct but similar-looking thermal expansion coefficient $\alpha=0$.\cite{LLElasticity,LLStatistical}  We consider an isotropic solid, for which $\kappa_{ij}=\kappa\delta_{ij}$ and $\beta_{ij}=\beta\delta_{ij}$, this $\beta$ not to be confused with the identical symbol sometimes used for the thermal expansion coefficient.\cite{LLFluid}

We also neglect the tensor viscosity $\eta_{iklm}$, which to lowest order in $k$ does not contribute to the modes.  The fluctuation of the tensor $\Pi_{ik}'$ \eqref{n-Pi} has a term from the viscosity $\sim \eta_{iklm} k_m v_l'$ and a term from the stress tensor $\sim \lambda'_{ik} \approx (\partial \lambda_{ik}/\partial w_{jl}) w_{jl}'$.  Then, by \eqref{n-udot}, $\lambda'_{ik} \sim w'_{jl} \sim k_j v'_l/\omega$.  Thus, for both propagating modes ($\omega \sim k$) and diffusive modes ($\omega \sim k^2$), the term in $\Pi_{ik}'$ due to viscosity is, at the least, of order $k$ relative to the term $\lambda_{ik}'$, and is therefore neglected in the long wavelength limit.

\subsection{Thermal Diffusion}
For the normal solid it is convenient to work with $\rho$ and $\sigma=s/\rho$ as variables, because $\sigma$ diffuses but does not flow, and therefore is nearly conserved.  To see this note that, to lowest order in deviations from equilibrium, eq.~\eqref{n-sdot} and \eqref{n-rhodot} yield
\begin{equation}
\partial_{t}\sigma'=\frac{1}{\rho}\partial_{i} \left(\frac{\kappa}{T}\partial_{i}T' \right)\approx\frac{\kappa}{T(\partial\sigma/\partial T)_\rho}\nabla^{2}\sigma', \label{s-sigmadot}
\end{equation}
where we have used $\alpha=0$.\cite{LLFootnote} 
 This equation describes entropy diffusion, with $\sigma' \neq 0$ and
\begin{equation}
\omega=-iD_{T}k^{2}, \qquad D_{T}=\frac{\kappa}{\rho T(\partial\sigma/\partial T)_\rho}. \label{n-DL}
\end{equation}
For this mode $u'_{i}=v'_{i}=\rho'=0$.  If $\alpha$ is small but non-zero the frequency will not change to lowest order in $\alpha$, but from the equations for $\rho$, $\vec{g}$, and $\vec{u}$ these quantities would develop amplitudes proportional to $\sigma'$ and $\alpha$, and thus have negligible amplitude as $\alpha\rightarrow0$.  We consider only the case where the effects of $\alpha$ can be neglected. 

\subsection{Elastic Modes}
We obtain the elastic modes by taking $\sigma' = 0$ and neglecting dissipative and nonlinear terms in \eqref{n-udot}-\eqref{n-gdot}.  Thus, eq.~\eqref{n-udot} gives $\dot{u}_i' = v_i'$.
In the remainder of this work, all thermodynamic derivatives are taken at constant $\sigma$, and derivatives with respect to $\rho$ are taken at constant $w_{ik}$ and vice-versa, unless otherwise specified.  Further, when derivatives with respect to a specific component of $w_{ik}$ are taken, the other components of $w_{ik}$ are held fixed.  Then by \eqref{n-Pi} and \eqref{n-g}, eqs.~\eqref{n-gdot} and \eqref{n-rhodot} become\cite{ALfootnote} 
\begin{align}
0&=\rho\ddot{u}'_{i}+\frac{\partial P}{\partial\rho}\partial_{i}\rho' +\frac{\partial P}{\partial w_{jl}}\partial_{i}w'_{jl} -\frac{\partial\lambda_{ik}}{\partial \rho}\partial_{k}\rho' -\frac{\partial\lambda_{ik}}{\partial w_{jl}}\partial_{k}w'_{jl}, \label{n-gdot-elastic}\\
0&=\dot{\rho}'+\rho\partial_{i}\dot{u}'_{i}. \label{n-rhodot-elastic}
\end{align}

Clearly, $\sigma'$ does not couple to the other variables.  On linearizing, eq.~\eqref{n-rhodot-elastic} gives $\rho'=-\rho\partial_{i}u'_{i}$, so with \eqref{wDEF}, eq.~\eqref{n-gdot-elastic} becomes
\begin{align}
&0= \rho\ddot{u}'_{i}-\rho\frac{\partial P}{\partial\rho}\partial_{i}\partial_{k}u'_{k}    + \frac{\partial P}{\partial w_{jl}}\partial_{i} \partial_j u'_l \notag\\
&\qquad \qquad \qquad \qquad + \rho \frac{\partial\lambda_{ik}}{\partial \rho}\partial_k \partial_j u'_j  -\frac{\partial\lambda_{ik}}{\partial w_{jl}}\partial_{k} \partial_j u'_l. \label{n-gdot-elastic2}
\end{align}
The second term gives the pure fluidlike (longitudinal) response, which occurs for $P \neq 0$ (e.g., an imperfect solid or a solid under $P_a$), and the fifth term gives the pure solidlike (longitudinal and transverse) response.  
 
Appendix~\ref{DerivativeAppendix} shows that, for uniform static $P_{a}$, certain quantities are isotropic.  This permits us to define
\begin{align}
\frac{\partial P}{\partial w_{jl}} \equiv \frac{\partial P}{\partial w} \delta_{jl}, \quad \frac{\partial \lambda_{jl}}{\partial \rho} \equiv \frac{\partial \lambda}{\partial \rho} \delta_{jl}, \quad \frac{\partial \lambda}{\partial w} \equiv K+ \frac{4}{3} \mu_V.
\label{AppDefs1}
\end{align}
Appendix~\ref{DerivativeAppendix} also shows that %The thermodynamic derivatives in \eqref{n-gdot-elastic2} are given in 
\begin{align}
\left. \frac{\partial \lambda_{ik}}{\partial w_{jl}} \right|_{\rho,\sigma} =&\frac{\partial \lambda}{\partial w} \delta_{ik} \delta_{jl} + \mu_V \left(\delta_{ij}{\delta_{kl}} + \delta_{kj}\delta_{il} - 2 \delta_{ik} \delta_{jl}\right).
\label{Appdlamdw}
\end{align}
%Substitution of \eqref{dPdwSYM}, \eqref{dlambdadwSYM} and \eqref{dlambdadrhoSYM} into \eqref{n-gdot-elastic2} gives 
Thus \eqref{n-gdot-elastic2} gives 
\begin{align}
0&\approx \rho\ddot{u}'_{i} - \rho \frac{\partial P}{\partial \rho} \partial_i \partial_k u'_k + \frac{\partial P}{\partial w} \partial_i \partial_k u_k' \notag\\
&+ \rho \frac{\partial \lambda}{\partial \rho} \partial_i \partial_k u_k' %-\left(K+\frac{1}{3}\mu_{V} \right)\partial_{i}\partial_{j}u'_{j} -\mu_{V}\nabla^{2}u'_{i}.
-\left(\frac{\partial \lambda}{\partial w}-\mu_{V} \right)\partial_{i}\partial_{j}u'_{j} -\mu_{V}\nabla^{2}u'_{i}.
\label{n-gdot-elastic2.5}
\end{align}
On letting $\partial_i \rightarrow i k_i$ and $\partial_t \rightarrow -i \omega$, eq.~\eqref{n-gdot-elastic2.5} becomes
\begin{align}
0&\approx (-\rho\omega^{2} + \mu_{V}) k^{2}u'_{i} \notag\\
&  +\left[\rho \frac{\partial P}{\partial \rho} - \frac{\partial P}{\partial w} - \rho \frac{\partial \lambda}{\partial \rho}  +\left(\frac{\partial \lambda}{\partial w}-\mu_{V} \right) \right] k_i (\vec{k}\cdot\vec{u})  .
\label{n-gdot-elastic3}
\end{align}

{\bf Longitudinal Mode:} 
If $\vec{k}\cdot\vec{u}\ne0$, then \eqref{n-gdot-elastic3} shows that $u_{i}$ is along $k_{i}$, so the mode is longitudinal.  Moreover, eq.~\eqref{n-gdot-elastic3} gives the normal mode frequencies
\begin{align}
\omega^{2} &=\left[\frac{\partial P}{\partial \rho} - \frac{1}{\rho} \frac{\partial P}{\partial w} - \frac{\partial \lambda}{\partial \rho}  +\frac{1}{\rho} \frac{\partial \lambda}{\partial w} \right]k^{2} \notag\\
&=\left[c^{2}_{lL}+c^{2}_{lS}\right]k^{2} \equiv c_l^2 k^2,
\label{n-omegalong}
\end{align}
where 
\begin{align}
c_{lL}^2 \equiv& \frac{\partial P}{\partial \rho} - \frac{\partial \lambda}{\partial \rho}, \qquad 
c_{lS}^2 \equiv \frac{1}{\rho} \frac{\partial \lambda}{\partial w}- \frac{1}{\rho} \frac{\partial P}{\partial w}.
\label{n-LSv}
\end{align}
The liquid-like velocity $c_{lL}$ contains thermodynamic derivatives with respect to the density $\rho$, and the solid-like velocity $c_{lS}$ contains thermodynamic derivatives with respect to the strain $w_{ik}$.  Eq.~\eqref{n-omegalong} gives a velocity for longitudinal sound that is similar to that found in Ref.~\onlinecite{MartinParodi}. 

Appendix~\ref{DerivativeAppendix} finds the four derivatives in \eqref{n-LSv} in terms of $P_a$, which  
to second order in $P_{a}/K$ give
%\eqref{dPdwSYMexpand}, \eqref{dPdrhoexpand}, \eqref{dlambdadwDEF} and \eqref{dlambdadrhoSYMexpand} give 
\begin{align}
c_{lL}^2 =& \frac{P_a}{\rho} \left(\frac{K^*}{K} - 1 \right) + \frac{P_a^2 K^*}{2 \rho K^2} \left(1 - \frac{K^*}{K} + \frac{\rho}{K^*} \frac{\partial K^*}{\partial \rho}  \right),\label{clLexpand} \\
c_{lS}^2 =& \frac{K+\frac{4}{3}\mu_{V}}{\rho}+ \frac{P_a}{\rho} \frac{K^*}{K} -  \frac{P_a^2 {K^*}^2}{2 \rho K^3},\label{clSexpand}
\end{align}
where $K^*$ is defined in \eqref{Kmustar}.  For $P_a=0$ we have %eq.~\eqref{n-omegalong} 
$c_{l}^{2}=c_{lL}^{2}+c_{lS}^{2}=[K+({4}/{3})\mu_{V}]/\rho$, which agrees with Ref.~\onlinecite{LLElasticity} for an ordinary solid. %, with Ref.~\onlinecite{YooDorsey} for an ordinary solid% under certain assumptions, \cite{YooFootnote} and with Ref.~\onlinecite{AL69} for a supersolid when $\rho_s$ is set to zero.\cite{ALFootnote2}}

{\bf Transverse Mode:}  If $\vec{k}\cdot\vec{u}=0$, so that the mode is transverse, then \eqref{n-gdot-elastic3} gives the normal mode frequencies
\begin{equation}
\omega^{2}=\frac{\mu_{V}}{\rho}k^{2}. \label{n-omegatran}
\end{equation}
From \eqref{n-rhodot-elastic}, for the transverse mode $\rho'=0$.  Eq.~\eqref{n-omegatran} agrees with Ref.~\onlinecite{LLElasticity} for an ordinary solid.

For both longitudinal and transverse mode frequencies, eq.~\eqref{s-sigmadot} is satisfied by $\sigma' = 0$.

\subsection{Lattice Diffusion}
The lattice diffusion mode is the most subtle of the modes.  For this mode, as for the elastic modes, we consider that $\sigma$ is constant, but we do not take $v'_{i}=\dot{u}'_{i}$.  Rather, we assume that $\omega=-iD_{L}k^{2}$, where the lattice mode diffusion constant $D_{L}>0$ is to be determined, and we keep the dissipative terms in the equations of motion for $v'_{i}$, $u'_{i}$, and $\rho'$.  

With $\beta_{ij}=\beta\delta_{ij}$ (i.e., an isotropic solid), $\alpha_{ij}=0$, and setting $\sigma' = 0$, eqs.\,\eqref{n-rhodot} and \eqref{n-udot} give
\begin{align}
-i\omega\rho' &=-\rho (ik_{i})v'_{i}, \label{n-rhodotD}\\
-i\omega u'_{i}&=v'_{i}+\beta(ik_{k})\lambda'_{ik}. \label{n-udotD}
\end{align}
If we assume that the mode is longitudinal, with $v_{i}'\sim k_{i}$ (the consistency of this assumption to be determined below), then the first of these equations implies that $v'_{i}\sim k_{i}\rho'$.  
Therefore in \eqref{n-gdot} the term $\partial_t g_i \sim \omega \rho v_i'$ is of order $k^2$ relative to the $k \rho'$ dependence of $\partial_k \Pi_{ik}'$, and is neglected in the long wavelength limit.
As a consequence, $\partial_k \Pi_{ik}' \approx 0$: the contributions from the liquid-like part $P'\delta_{ik}$ and from the solid-like part $-\lambda'_{ik}$ nearly cancel.  This can only occur for an imperfect solid or a solid under applied pressure $P_a$, which has both liquid-like and solid-like responses. 

Thus, neglecting the $\partial_t g_i \sim \omega \rho v_i'$ term and neglecting the viscosity $\eta_{iklm}$ (as discussed above), eq.~\eqref{n-gdot} gives
\begin{equation}
ik_{i}\rho'  \frac{\partial P}{\partial\rho} - i k_{k} \rho' \frac{\partial \lambda_{ik}}{\partial \rho}=-k_{k}k_{j}\frac{\partial\lambda_{ik}}{\partial w_{jl}}u'_{l}  + \frac{\partial P}{\partial w_{jl}} k_{i} k_{j} u_l' . \label{n-rhouD0}
\end{equation}
Substitution from \eqref{AppDefs1} and \eqref{Appdlamdw} gives
 \begin{align}
 &ik_{i}\rho'  \frac{\partial P}{\partial\rho} - i k_{i} \rho' \frac{\partial \lambda}{\partial \rho} \notag\\
 &\quad =-\left(\frac{\partial \lambda}{\partial w}-\mu_{V} \right) k_{i}k_{l}u'_{l}  - \mu_V k^2 u'_{i} + \frac{\partial P}{\partial w} k_{i} k_{l} u_l' .
\label{n-rhouD}
\end{align}
All but one term in \eqref{n-rhouD} is along $k_{i}$, and the remaining term is along $u'_{i}$.  Therefore we deduce that $u'_{i}$ is along $k_{i}$, and thus $k_{i}k_{l}u'_{l}=k^{2}u'_{i}$.  Then \eqref{n-rhouD} becomes
\begin{align}
& i \rho' \left(  \frac{\partial P}{\partial\rho} - \frac{\partial \lambda}{\partial \rho} \right) =   -k_{l}u'_{l}\left(\frac{\partial \lambda}{\partial w} -  \frac{\partial P}{\partial w} \right). \label{n-rhouD3}
\end{align}

Further, eq.~\eqref{n-udotD} gives, 
on taking $\lambda'_{ik} = (\partial \lambda_{ik}/\partial w_{jl}) w_{jl}' + (\partial \lambda_{ik}/\partial \rho) \rho'$, 
and taking $u_i'$ along $k_i$, 
\begin{align}
&\left[-i\omega + \beta \frac{\partial \lambda}{\partial w}  k^{2} \right]u'_{i} =v'_{i} + i k_i \beta  \frac{\partial \lambda}{\partial \rho} \rho' . 
\label{n-urhovD2}
\end{align}
Since $u_i'$ is along $k_i$, eq.~\eqref{n-urhovD2} implies $v_i'$ also along $k_i$.  Hence the mode is longitudinal.

We now use \eqref{n-rhodotD} and the sound velocities of \eqref{n-LSv} to eliminate $\rho'$ from \eqref{n-rhouD3} and \eqref{n-urhovD2}.  Then \eqref{n-rhouD3} multiplied by $\omega$ gives
\begin{equation}
i\rho k_{i}v'_{i} c_{lL}^2 =-\omega k_{i}u'_{i} \rho c_{lS}^2,
\label{v-u}
\end{equation}
and \eqref{n-urhovD2} multiplied by $\omega$ gives
\begin{align}
& \left[-i\omega + \beta \frac{\partial \lambda}{\partial w}  k^2 \right]\omega u'_{i}  =\left[ \omega + i \beta \rho \frac{\partial \lambda}{\partial \rho}  k^2 \right] v'_{i} .
\label{v-u2}
\end{align}

Since $u'_{i}$ and $v'_{i}$ are along $k_{i}$, eq.~\eqref{v-u} implies that for the diffusive mode 
\begin{align}
v_{i}'= i\omega u_{i}'(c^{2}_{lS}/c^{2}_{lL})=-\dot{u}_{i}'(c^{2}_{lS}/c^{2}_{lL}), 
\end{align}
which is independent of $\omega$.  We interpret this as the lattice velocity $\dot{u}_{i}'$ being out of phase relative to the matter velocity $v_{i}'$ so that the fluid and lattice stresses cancel.  

Combining \eqref{v-u} and \eqref{v-u2} then yields 
\begin{align}
\omega \left( c_{lS}^2+c_{lL}^2\right)
%=&-i k^2 \beta \rho \left[c_{lS}^2 \frac{\partial P}{\partial \rho} + c_{lL}^2 \frac{1}{\rho} \frac{\partial P}{\partial w} \right] \notag\\
=&-i k^2 \beta \rho \left[c_{lS}^2 \frac{\partial \lambda}{\partial \rho} + c_{lL}^2 \frac{1}{\rho} \frac{\partial \lambda}{\partial w} \right] \notag\\
=&-i k^2 \beta \rho \left[ \frac{\partial \lambda}{\partial w}\frac{\partial P}{\partial \rho} - \frac{\partial \lambda}{\partial \rho} \frac{\partial P}{\partial w} \right].
%=& -i k^2 \beta \rho \left[ c_{lS}^2 c_{lL}^2 + c_{lS}^2 \frac{\partial \lambda}{\partial \rho} + c_{lL}^2 \frac{1}{\rho} \frac{\partial P}{\partial w}\right] \notag\\
%=& -i k^2 \beta \rho \left[c_{lS}^2 \frac{\partial P}{\partial \rho} + c_{lL}^2 \frac{1}{\rho} \frac{\partial \lambda}{\partial w} - c_{lS}^2 c_{lL}^2\right] . \label{n-omegaD2}
\end{align}
Therefore
\begin{align}
D_{L} =i\frac{\omega}{k^{2}} %=& \beta \rho \left[ \frac{c_{lS}^2 \frac{\partial \lambda}{\partial \rho} + c_{lL}^2 \frac{1}{\rho} \frac{\partial \lambda}{\partial w}}{c_{lS}^{2}+c_{lL}^{2}}\right] \notag\\
=& \beta \rho \left[\frac{ \frac{\partial \lambda}{\partial w}\frac{\partial P}{\partial \rho} - \frac{\partial \lambda}{\partial \rho} \frac{\partial P}{\partial w} }{ \frac{\partial }{\partial w} (\lambda - P) -  \rho \frac{\partial }{\partial \rho} (\lambda - P)}\right].
\label{n-D_{L}}
\end{align}
%\textbf{Under certain assumptions,\cite{YooFootnote} 
For $P_a = 0$, eq.~\eqref{n-D_{L}} agrees with Ref.~\onlinecite{YooDorsey}\cite{YooFootnote} and with Ref.~\onlinecite{Zippelius}.\cite{ZippeliusFootnote}   For either a pure liquid or a pure solid, $D_{L}\rightarrow0$:   

\begin{itemize}
\item{}For a pure liquid, derivatives with respect to strain go to zero: $\partial \lambda/\partial w \rightarrow 0$ and $\partial P/\partial w \rightarrow 0$.  Therefore $D_L \rightarrow 0$.
\item{}For a pure solid, derivatives with respect to density (at constant strain) go to zero: $\partial \lambda/\partial \rho \rightarrow 0$ and $\partial P/\partial \rho \rightarrow 0$.  Therefore $D_L \rightarrow 0$.
\end{itemize}

If the system is not supersolid, and if the samples are not perfect, then it is consistent to interpret the observations of Ref.~\onlinecite{RitRep09} in terms of this lattice diffusion mode.  

Substitution for the four derivatives in \eqref{n-D_{L}} from Appendix~\ref{DerivativeAppendix} %of \eqref{dPdwSYMexpand}, \eqref{dPdrhoexpand}, \eqref{dlambdadwDEF} and \eqref{dlambdadrhoSYMexpand} into \eqref{n-D_{L}} 
gives, to lowest order in $P_{a}/K$,
\begin{align}
&D_{L} = \notag\\
&\frac{\beta V P_a^2}{K^2} \left[ \frac{V}{2} \frac{\partial^2 K}{\partial V^2} + \frac{K}{K+\frac{4}{3}\mu_V} \frac{\partial K}{\partial V} - \frac{V}{K+\frac{4}{3}\mu_V} \left( \frac{\partial K}{\partial V} \right)^2 \right],
\label{DLexpand}
\end{align}
where derivatives with respect to $V$ are taken at constant $(w_{ik},\sigma,N)$.  The form \eqref{DLexpand} does not apply to case of a pure liquid (whereas \eqref{n-D_{L}} is general), because it does not permit $P$ to have terms {\it independent} of strain.  Recall that we have assumed that it 
is valid to expand $K$ around $P_{a}=0$.  If, in $D_{L}$, all other dependences on $P_{a}$ can be neglected, then \eqref{DLexpand} implies that $D_{L}\sim P_{a}^{2}$.  

\section{Longitudinal Response of Normal Solid}
\label{ModeGeneration}
Recall that $\beta$ has units of $D$ divided by pressure.  As $T\rightarrow 0$ we expect that, by the Arrhenius equation, $D\rightarrow 0$ as $\exp[-\Delta/k_{B}T]$, where $\Delta$ is a hopping energy, because the hopping rate should yield such a dependence.  Therefore, if the wavevector $k$ is replaced by $d^{-1}$, where $d$ is a characteristic distance (the plate separation in Ref.~\onlinecite{RitRep09}), then the characteristic response time $\tau\sim\omega^{-1}\sim (Dk^{2})^{-1}\sim d^{2}/D$.  Hence the view that the experimental results of Ref.~\onlinecite{RitRep09} are due to a lattice diffusion mode leads to the conclusion that $\tau$ varies as $\exp[\Delta/k_{B}T]$.  Indeed, such a dependence is observed, with $\Delta\sim30$~mK.  It would be useful to test for the predicted $d^{2}$-dependence. For instance, the present theory predicts that changing the plate separation in the pancake cell of Ref.~\onlinecite{RitRep09} from 100~$\mu$m to 200~$\mu$m should yield a relaxation time approximately four times longer.

This mode provides a means for vacancy flow to equilibrate vacancy concentrations.  It is consistent with the observation of Ref.~\onlinecite{RitRep09} that pressure decreases during an anneal, and when the system relaxes at constant temperature.  We interpret this to mean that vacancies diffuse to or from the surface.

We now turn to how a normal solid will respond to the two devices usually employed to generate a disturbance: a heater and a transducer.  Since there are three longitudinal modes (thermal diffusion, lattice diffusion, and elastic waves), it would appear that there is need for an additional independent generator.  Perhaps surface properties introduce a new boundary condition that amounts to having an independent generator.  For example, the material against the solid $^4$He may cause the $^4$He surface to prefer vacancies, as opposed to atoms.   Thus the surface treatment may affect the behavior of both heaters and transducers.  This argument applies to any two ordinary solids, and there may be some for which this can be readily tested.  Hence two macroscopically identical heaters or transducers made of different materials, or of the same material but with different surface treatment, would not show identical behavior.  Since $v_{i}-\dot{u}_{i}\approx 0$ for the temperature mode and the elastic modes, one way to characterize the response of a surface is in terms of $v_{i}-\dot{u}_{i}$.  Thus $(v_{i}-\dot{u}_{i})/P'$ for a longitudinally  moving transducer  and $(v_{i}-\dot{u}_{i})/T'$ for a heater would characterize differences in the response to different surface conditions, and the extent to which they can generate the lattice diffusion mode. 

\section{Summary and Conclusions}
\label{Summary}
We have applied the Andreev-Lifshitz theory of supersolid dynamics to an ordinary solid with lattice defects -- specifically, with vacancies in mind.  At the thermodynamic level, this theory includes an internal pressure $P$, distinct from the applied pressure $P_a$ and the stress tensor $\lambda_{ik}$.  For the Andreev-Lifshitz theory this is necessary to permit a continuous variation from a supersolid to a superfluid.  Under uniform static $P_{a}$, we have $\lambda_{ik}=(P-P_{a})\delta_{ik}$.  For $P_{a}\ne0$, Maxwell relations imply that $P\sim P_{a}^{2}$.  These results are not conventional; Ref.~\onlinecite{LLElasticity} does not include $V$ as a distinct extensive thermodynamic variable, nor its thermodynamically conjugate variable $P$.  In the present work many derivatives involving $V$ are at fixed strain $w_{ik}$, which is also unconventional, since normally one assumes that $\delta w_{ii} = -\delta \rho/\rho$.\cite{LLElasticity}  Nevertheless, the variables of Andreev and Lifshitz must be taken if vacancies are to be permitted.  

For an isotropic model, the normal modes were obtained.  There are, as expected, two sets of propagating transverse modes, with velocities as expected.  There also are, as expected, a set of propagating longitudinal modes, but with velocities containing both solid-like and liquid-like contributions, and which depend upon $P_{a}$.  In addition there are two diffusive longitudinal modes: a well-known mode that dominantly involves temperature, and another mode involving lattice defects (i.e., vacancies).  Our analysis of the physical nature of this mode shows that it is surprisingly complex.  It involves the mass density $\rho$, the lattice velocity $\dot{u}_{i}$, and the mass-flow velocity $v_{i}$, with the fluid-like pressure $P$ associated with $\rho$ essentially canceling the solid-like stress $\lambda$ associated with $u_{i}$.  

In a separate work\cite{SearsSasALSS10} we discuss the normal modes of the full Andreev and Lifshitz theory for a supersolid, which has nine variables.  As Ref.~\onlinecite{AL69} established at $T=0$, there are four pairs of propagating modes.  Three pairs are essentially the elastic modes we have studied here, with a weak coupling to the superfluid.  The fourth pair is basically a fourth sound mode, where the normal fluid is entrained by the lattice.  These propagating modes, in the presence of a finite $P_{a}$, and their generation by transducers and heaters, have been considered in Ref.\onlinecite{SearsSasSSGen}.  We also find a rather complex additional mode, not considered in Ref.~\onlinecite{AL69}, which is diffusive.\cite{SearsSasALSS10}  Although the additional supersolid diffusive mode is similar to the normal solid diffusive mode found in the present work (e.g., zero net stress, and distinct mass and lattice motion), %both its diffusion constant and 
its mode structure differs significantly.  %We find the supersolid diffusive mode frequency to be similar to that of ordinary thermal diffusion, and that the mode structure is frequency-independent.  
The supersolid diffusive mode is characterized by three velocities: ${v'_n}_i$, ${v'_s}_i$, and $\dot{u}'_i$, associated respectively with the normal mass, superfluid mass, and the lattice.  For supersolid $^4$He with $P_a \ll K$, we find that ${v'_s}_i \gg {v'_n}_i \gg \dot{u}'_i$.  We also find that $g' = \rho_n v_n' + \rho_s v_s' \approx 0$.  If $^4$He is a genuine supersolid, then this mode provides an alternate explanation for the exponential time-dependence of the pressure decay observed by Ref.~\onlinecite{RitRep09}. 

We close with the following comment.  Ref.~\onlinecite{AL69} predicted that supersolidity will occur because of quantum diffusion, a situation that occurs at such low temperatures that the relevant bulk diffusion processes are temperature-independent.  Ref.~\onlinecite{RitRep09} observe temperature-dependent relaxation; therefore their system is not at a low enough temperature to be in the quantum diffusive regime.  Note that quantum spin tunneling is an established phenomenon, wherein the magnetic relaxation rate saturates at low enough temperatures.\cite{Chudnovsky,Tejada,ChudTejBook} 

\section{Acknowledgements}
This work was partially supported by the Department of Energy through grant DE-FG02-06ER46278.

\appendix

\section{Andreev-Lifshitz Supersolid}
\label{ALAppendix}

\subsection{Thermodynamics}
Consider a general frame of reference, with non-zero superfluid velocity $\vec{v}_{s}$ and normal fluid velocity $\vec{v}_{n}$.  Let $\vec{u}$ be the local displacement of the crystal sites relative to their equilibrium, and take the strain to be given by $w_{ik}=\partial_{i}u_{k}$.  Then by thermodynamics the differential of the energy density $\epsilon$ is given by
\begin{equation}
d\epsilon=Tds+\lambda_{ik}dw_{ik}+ \mu d\rho +\vec{j}_{s}\cdot d\vec{v}_{s}+\vec{v}_{n}\cdot d\vec{g}. 
\label{thermodiffB}
\end{equation}
Here $\lambda_{ik}$ is an elastic tensor density (with the same units as pressure $P$), $\mu$ is the chemical potential (with units of velocity squared), $\vec{j}_{s}=\vec{g}-\rho\vec{v}_{n}$ (a requirement of Galilean relativity), $\rho = \rho_n + \rho_s$ (the sum of the normal and superfluid densities), and $\vec{g} = \rho_n \vec{v}_n + \rho_s \vec{v}_s$.  By thermodynamic extensivity we also have 
\begin{equation}
\epsilon=-P+Ts+\lambda_{ik}w_{ik}+\mu \rho +\vec{j}_{s}\cdot\vec{v}_{s}+\vec{v}_{n}\cdot\vec{g}
\label{thermo}
\end{equation}
and the Gibbs-Duhem relation
\begin{equation}
0=-dP+sdT+w_{ik}d\lambda_{ik}+\rho d\mu +\vec{v}_{s}\cdot d\vec{j}_{s}+ \vec{g}\cdot d\vec{v}_{n}.
\label{Gibbs-Duhem}
\end{equation}
The system will be in equilibrium when the thermodynamic forces $\partial_{i}T$, $\partial_{i}\lambda_{ik}$, $\partial_i \mu$, $\partial_{i}{v_{n}}_{j}$, and $\partial_{i}{j_{s}}_{i}$ are all zero.  

\subsection{Dynamics}
The thermodynamic variables $\epsilon$, $s$, $u_{i}$, $\rho$, $\vec{v}_{s}$, and $\vec{g}$ are taken to satisfy equations of motion that are first order in time and that satisfy appropriate properties under space rotation and inversion, and under time-reversal.  Thus $\epsilon$, $\rho$, and $\vec{g}$ satisfy conservation laws (a flux but no source), the phase gradient $\vec{v}_{s}$ is proportional to a gradient (a type of flux, with no source), and the displacement $u_{i}$ has a source but no flux.  Thus
\begin{eqnarray}
\partial_{t}\epsilon+\partial_{i}Q_{i}&=&0, \label{s-epsilondot}\\
\partial_{t}s+\partial_{i}f_{i}&=&\frac{R}{T}, \label{s-sdot}\quad (R\ge0), \\
\partial_{t}u_{i}&=&U_{i}, \label{s-udot}\\
\partial_{t}\vec{v}_{s}+\vec{\nabla}\theta&=&0, \label{s-vsdot}\\
\partial_{t}\rho+\partial_{i}g_{i}&=&0, \label{s-rhodot}\\
\partial_{t}g_{i}+\partial_{k}\Pi_{ik}&=&0. \label{s-gdot}
\label{eqnsmot}
\end{eqnarray}
(The source $U_{i}$ was implicit in previous theories.\cite{AL69,Saslow77})  The unknown fluxes $Q_{i}$, $f_{i}$, $\phi$, and $\Pi_{ik}$, and the unknown sources $R$ and $U_{i}$, are determined by subjecting them to the condition that, when applied to the thermodynamic equation (\ref{thermodiffB}), the density $R$ of the rate of dissipated energy be non-negative.  Note that $g_{i}$ is already known, and $Q_{i}$ and $R$ will not be needed.  For $f_{i}$, $U_{i}$, $\theta$, and $\Pi_{ik}$ we have, when terms non-linear in velocities and strains are neglected, 
\begin{align}
f_{i}&=sv_{ni}-\frac{\kappa_{ij}}{T}\partial_{j}T-\frac{\alpha_{ij}}{T}\partial_{l}\lambda_{lk}, \label{s-f}\\
U_{i}&=v_{ni}+\frac{\alpha_{ij}}{T}\partial_{j}T+\beta_{ij}\partial_{l}\lambda_{lk}, \label{s-U}\\
\theta &=\mu-\zeta_{ik}\partial_{k}v_{ni}-\chi\partial_{k}j_{sk}, \label{s-theta}\\
\Pi_{ik}&=(P\delta_{ik}-\lambda_{ik})-\eta_{iklm}\partial_{m}v_{nl}-\zeta_{ik}\partial_{l}j_{sl}. \label{s-Pi}
%\label{fluxsource}
\end{align}
In each of these equations, the last two terms are dissipative and the preceding terms are reactive.  

Refs.~\onlinecite{AL69} and \onlinecite{Saslow77} obtain a nonlinear term in $U_{i}$, which may be obtained by letting $v_{i}\rightarrow v_{i}-v_{j}\partial_{j}u_{i}$.  On the other hand, Ref.~\onlinecite{YooDorsey} obtains two nonlinear terms, which may be obtained by letting $v_{i}\rightarrow v_{i}-v_{j}\partial_{j}u_{i}-u_{i}\partial_{j}v_{j}$.\cite{YooDorseyFootnote}  %The last term $-u_{i}\partial_{j}v_{j}$, proportional to the lattice position $u_{i}$, would cause $\dot{u}_{i}$ to depend upon the choice of origin, and therefore is not translationally invariant. %See (4.14-4.6) of Saslow's PRB15, where we did not employ an explicit name for $U_{i}$, simply using $-\dot{u}_{i}$, and we used $f_{i}=sv_{ni}+(q_{i}/T)$.  That work also computed $Q_{i}$ and $R$ in detail, which we do not do here. 

\section{Relevant Thermodynamic Derivatives}
\label{DerivativeAppendix}
%Let $\sigma = s/\rho$ be the entropy per unit mass.  \frac{\partial P}{\partial w_{jl}} 
%In a later section we show that the four quantities $(\partial P/\partial w_{jl})_{\rho,\sigma}$, $(\partial P/\partial \rho)_{w_{ik},\sigma}$, $(\partial \lambda_{ik}/\partial w_{jl})_{\rho,\sigma}$, and $(\partial \lambda_{ik}/\partial \rho)_{w_{ik},\sigma}$ are needed to determine both the ordinary elastic modes as well as the lattice diffusion mode.  
In what follows, the quantities $(\partial P/\partial w_{jl})_{\rho,\sigma}$, $(\partial P/\partial \rho)_{w_{ik},\sigma}$, $(\partial \lambda_{ik}/\partial w_{jl})_{\rho,\sigma}$, and $(\partial \lambda_{ik}/\partial \rho)_{w_{ik},\sigma}$ are obtained in terms of $P_{a}$ and the elastic constants.

{(1)}
With ${w_{11}^{(0)}} = {w_{22}^{(0)}} = {w_{33}^{(0)}}$, eq.~\eqref{dPdw1.5} gives
\begin{align}
\left. \frac{\partial P}{\partial w_{ik}} \right|_{V,s,\rho}=&  3 K^* \delta_{ik} w_{11}^{(0)} \equiv \frac{\partial P}{\partial w} \delta_{ik},
\label{dPdwSYM}
\end{align}
where $\partial P/\partial w$ is defined for later convenience.
Substitution for $w_{11}^{(0)}$ from \eqref{w11expand} gives, to second order in $P_a/K$,
\begin{align}
\frac{\partial P}{\partial w}  \approx -P_a \frac{K^*}{K} + \frac{P_a^2 {K^*}^2}{2K^3}. 
\label{dPdwSYMexpand}
\end{align}

{(2)}
From \eqref{P} we have 
\begin{align}
\left. \frac{\partial P}{\partial \rho} \right|_{\sigma, w_{ik}} = \frac{9}{2}{ w_{11}^{(0)}}^2 \left. \frac{\partial K^*}{\partial \rho} \right|_{\sigma, w_{ik}}.
\label{dPdrho}
\end{align}
By \eqref{Kmustar},
\begin{align}
\left. \frac{\partial K^*}{\partial \rho} \right|_{\sigma, w_{ik}} =& \left[ \frac{\partial }{\partial \rho} \left(K - V  \left. \frac{\partial K}{\partial V}\right|_{\sigma, w_{ik},N} \right) \right]_{\sigma, w_{ik}} \notag\\
=& \frac{V^2}{\rho} \left. \frac{\partial^2 K}{\partial V^2}\right|_{\sigma, w_{ik},N} .
\end{align}
Thus \eqref{dPdrho} can be written as
\begin{align}
\left. \frac{\partial P}{\partial \rho} \right|_{\sigma, w_{ik}} = \frac{9V^2}{2\rho} { w_{11}^{(0)}}^2 \left. \frac{\partial^2 K}{\partial V^2}\right|_{\sigma, w_{ik},N}.
\label{dPdrho2}
\end{align}

To second order in $P_a/K$, eqs.~\eqref{w11expand}, \eqref{dPdrho} and \eqref{dPdrho2} give
\begin{align}
\left. \frac{\partial P}{\partial \rho} \right|_{\sigma, w_{ik}} \approx \frac{1}{2}\frac{P_a^2}{K^2} \left. \frac{\partial K^*}{\partial \rho} \right|_{\sigma, w_{ik}} = \frac{V^2 P_a^2}{2\rho K^2}  \left. \frac{\partial^2 K}{\partial V^2}\right|_{\sigma, w_{ik},N}.
\label{dPdrhoexpand}
\end{align}

(3) 
From \eqref{lambdaNEW} we have 
\begin{align}
\left. \frac{\partial \lambda_{ik}}{\partial w_{jl}} \right|_{\rho,\sigma} =&\left(K - \frac{2}{3}\mu_V \right) \delta_{ik} \delta_{jl} + \mu_V \left(\delta_{ij}{\delta_{kl}} + \delta_{kj}\delta_{il}\right).
\label{dlambdadw}
\end{align}
We now define
\begin{align}
\frac{\partial \lambda}{\partial w} \equiv K+\frac{4}{3} \mu_V,
\label{dlambdadwDEF}
\end{align}
so that 
\begin{align}
\left. \frac{\partial \lambda_{ik}}{\partial w_{jl}} \right|_{\rho,\sigma} =&\frac{\partial \lambda}{\partial w} \delta_{ik} \delta_{jl} + \mu_V \left(\delta_{ij}{\delta_{kl}} + \delta_{kj}\delta_{il} - 2 \delta_{ik} \delta_{jl}\right).
\label{dlambdadwSYM}
\end{align}

(4)
From \eqref{lambdaNEW} we also have 
\begin{align}
\left. \frac{\partial \lambda_{ik}}{\partial \rho} \right|_{w_{ik},\sigma} =& \left(\left. \frac{\partial K}{\partial \rho} \right|_{w_{ik},\sigma} - \frac{2}{3}\left. \frac{\partial \mu_V}{\partial \rho} \right|_{w_{ik},\sigma} \right) \delta_{ik} w_{ll}^{(0)} \notag\\
&+ \left. \frac{\partial \mu_V}{\partial \rho} \right|_{w_{ik},\sigma} \left(w_{ik}^{(0)} + w_{ki}^{(0)}\right).
\label{dlambdadrho}
\end{align}
With 
\begin{align}
\frac{\partial K}{\partial \rho}_{w_{ik},\sigma} = - \frac{V}{\rho} \frac{\partial K}{\partial V}_{w_{ik},\sigma,N} = \frac{K^*-K}{\rho},
\end{align}
and with a similar relation for $\mu_V$, eq.~\eqref{dlambdadrho} gives
\begin{align}
\left. \frac{\partial \lambda_{ik}}{\partial \rho} \right|_{w_{ik},\sigma} =& \left( \frac{K^* - K}{\rho} - \frac{2}{3}\frac{\mu_V^* - \mu_V}{\rho} \right) \delta_{ik} w_{ll}^{(0)} \notag\\
&+\frac{\mu_V^* - \mu_V}{\rho} \left(w_{ik}^{(0)} + w_{ki}^{(0)}\right).
\label{dlambdadrho1}
\end{align}

With \eqref{wik0SYM} and ${w_{11}^{(0)}} = {w_{22}^{(0)}} = {w_{33}^{(0)}}$,
\begin{align}
\left. \frac{\partial \lambda_{ik}}{\partial \rho} \right|_{w_{ik},\sigma} =& \left( \frac{K^* - K}{\rho}  \right) \delta_{ik} w_{ll}^{(0)} = 3 \left( \frac{K^* - K}{\rho}  \right) \delta_{ik} w_{11}^{(0)}  \notag\\
\equiv& \frac{\partial \lambda}{\partial \rho} \delta_{ik},
\label{dlambdadrhoSYM}
\end{align}
where $\partial \lambda/\partial \rho$ is defined for later convenience.
To second  order in $P_a/K$, eq.~\eqref{w11expand} gives
\begin{align}
\frac{\partial \lambda}{\partial \rho} \approx  \Big(1-\frac{K^*}{K}\Big) \left[\frac{P_a}{\rho} - \frac{P_a^2 K^*}{2 \rho K^2} \right].
\label{dlambdadrhoSYMexpand}
\end{align}


\begin{thebibliography}{}
\bibitem{Thouless69} D. J. Thouless, Ann. Phys. (N.Y.) 52, 403 (1969).  This contains the remark that, for a lattice of bosons, vacancies could be ``in the lowest Bloch state with a finite probability, so the system would be `super' but not `fluid'\,''. %The flow of a dense superfluid
\bibitem{NotationFootnote} The present work uses the notation of Ref.~\onlinecite{Saslow77}, which follows Ref.~\onlinecite{AL69} for the dissipative coefficients.  On the other hand, Ref.~\onlinecite{YooDorsey} seems to use the notation of Ref.~\onlinecite{Zippelius} for the dissipative coefficients, but follows Ref.~\onlinecite{AL69} in using $\vec{j}$ for the momentum density.  The present work, Ref.~\onlinecite{FC76}, Ref.~\onlinecite{Zippelius}, and Ref.~\onlinecite{MartinParodi} use $\vec{g}$ for the momentum density.
\bibitem{DefectFootnote} If $N_{L}$ is the number of lattice sites and $N_{os}$ is the number of on-site atoms, then $N_{L}=N_{os}+N_{V}$.  In addition, if there are $N_{i}$ interstitial atoms, then $N=N_{os}+N_{i}$.  With the energy differential taking the form $dE=\dots+\phi_{L}dN_{L}+\phi_{os}dN_{os}+\phi_{i}dN_{i}$, we then have $dE=\dots+\phi_{L}dN_{V}+(\phi_{L}+\phi_{os})dN_{os}+\phi_{i}dN_{i}$.  If the on-site and interstitials are in equilibrium, then $\phi_{L}+\phi_{os}=\phi_{i}$.  Further, if the vacancies (subject to no conservation law) are in equilibrium, then $\phi_{V}=0$.  If both the vacancies and the on-site and interstitial atoms are in equilibrium, then both $\phi_{V}=0$ and $\phi_{os}=\phi_{i}$.  We do not consider interstitials and only consider the case where $P_{a}$ dominates the effect on $P$.
\bibitem{NonzeroPFootnote} In typical treatments of elasticity in solids it is implicit that $P=0$, even for $P_{a} \neq 0$.  To produce $P=0$, even for $w_{ll}^{(0)} \neq 0$, the present application of AL theory requires $K^*=0$.  For that to hold, by  eq.~\eqref{Kmustar} $K$ must vary linearly with $V$ at fixed strain $w_{ik}$; this appears to be unlikely.
\bibitem{ALfootnote}  In the linear approximation the corresponding two equations of Ref.~\onlinecite{AL69} (see its eq.(19)), on taking $\rho_s \rightarrow 0$, agree with the above two equations.
\bibitem{YooFootnote}  To obtain agreement with Ref.~\onlinecite{YooDorsey}, we drop higher-order terms in the velocity, take $T=0$ (as in the present work), and neglect the static strain in eq.~\eqref{n-GibbsDuhem}, which gives $dP\approx \rho d \mu$.  Then $c_l^2$ of \eqref{n-omegalong} agrees with $c_{NS}^2$ found in Ref.~\onlinecite{YooDorsey}, and $D_L$ of \eqref{n-D_{L}} agrees with $D_2$ found in Ref.~\onlinecite{YooDorsey}.  However, for finite $P_a$ it is inconsistent to neglect the static strain (i.e., $\rho (\partial \mu/\partial \rho) \sim (P_a/K)^2 \sim w_{ik}^{(0)} (\partial \lambda_{ik}/\partial \rho)$).  %\textbf{Further, $\gamma$ of Ref.~\onlinecite{YooDorsey} is the same as $\partial \lambda/\partial \rho$ of the present work, which is $\sim P_a$.  Hence $\gamma=0$ in (what for solid $^4$He is) the inaccessible limit of $P_a=0$ taken by Ref.~\onlinecite{YooDorsey}, and should be neglected in their $D_2$.  Thus, for nonzero $P_a$ (as for solid $^4$He) the longitudinal sound velocity is $c_l$ rather than $c_{NS}$ and the diffusion constant for the lattice diffusion mode is $D_L$ rather than $D_2$.}  %Note that Ref.~\onlinecite{YooDorsey} seems to use the notation of Ref.~\onlinecite{Zippelius} for the dissipative coefficients, but follows Ref.\onlinecite{AL69} in using $\vec{j}$ for the momentum density.  The present work, Ref.\onlinecite{FC76}, Ref.\onlinecite{Zippelius}, and Ref.\onlinecite{MartinParodi} use $\vec{g}$ for the momentum density.
\bibitem{ZippeliusFootnote} To obtain agreement with Ref.~\onlinecite{Zippelius}, we take $\partial\lambda/\partial w$ to dominate the denominator of \eqref{n-D_{L}} and make the identifications, valid for zero static strain (such as $P_a = 0$ and vacancies in equilibrium),  that $\rho(\partial\lambda/\partial\rho)\rightarrow\gamma_{R}$ and $\rho(\partial P/\partial\rho)\rightarrow \chi_{R}^{-1}$.
\bibitem{YooDorseyFootnote} The term $-u_{i}\partial_{j}v_{j}$, proportional to the lattice position $u_{i}$, would cause $\dot{u}_{i}$ to depend upon the choice of origin; this is not translationally invariant.
\end{thebibliography}
\end{document}